# Affordable, rapid bootstrapping of space industry and solar system civilization

Philip T. Metzger, Ph.D., A.M. ASCE[1], Anthony Muscatello, Ph.D.[2], Robert P. Mueller, A.M. ASCE[3], and James Mantovani, Ph.D.[4]

[1] Physicist, Granular Mechanics and Regolith Operations Lab, NASA Kennedy Space Center, Philip.T.Metzger@nasa.gov, NE-S-1, Kennedy Space Center, FL 32899

[2] Chemist, Applied Chemistry Lab, NASA Kennedy Space Center, Anthony.C.Muscatello@nasa.gov, NE-S-2, Kennedy Space Center, FL 32899

[3] Aerospace Engineer, Surface Systems Office, NASA Kennedy Space Center, Rob.Mueller@nasa.gov, NE-S, Kennedy Space Center, FL 32899

[4] Physicist, Granular Mechanics and Regolith Operations Lab, NASA Kennedy Space Center, James.G.Mantovani@nasa.gov, NE-S-1, Kennedy Space Center, FL 32899

Abstract:

Advances in robotics and additive manufacturing have become game-changing for the prospects of space industry. It has become feasible to bootstrap a self-sustaining, self-expanding industry at reasonably low cost. Simple modeling was developed to identify the main parameters of successful bootstrapping. This indicates that bootstrapping can be achieved with as little as 12 metric tons (MT) landed on the Moon during a period of about 20 years. The equipment will be teleoperated and then transitioned to full autonomy so the industry can spread to the asteroid belt and beyond. The strategy begins with a sub-replicating system and evolves it toward full self-sustainability (full closure) via an in situ technology spiral. The industry grows exponentially due to the free real estate, energy, and material resources of space. The mass of industrial assets at the end of bootstrapping will be 156 MT with 60 humanoid robots, or as high as 40,000 MT with as many as 100,000 humanoid robots if faster manufacturing is supported by launching a total of 41 MT to the Moon. Within another few decades with no further investment, it can have millions of times the industrial capacity of the United States. Modeling over wide parameter ranges indicates this is reasonable, but further analysis is needed. This industry promises to revolutionize the human condition.

## Introduction

The solar system's resources are the key to humanity's future. Our civilization's demand for energy and material resources is rapidly growing toward the limits of the planet. There is mounting evidence that we are beginning to feel those limits in some of the non-renewable energy and mineral resources (Bentley 2002; de Almeida and Silva 2009; Lin and Liu 2010; Mudd and Ward 2008), and that they cannot support our current rates of population growth with industrialization for another century. Fortunately, the processes that formed our habitable Earth also endowed the solar system with literally billions of times more resources than exist on one planet alone (Hartmann 1985; Lewis and Lewis 1987; Lissauer 1993; Duke et al. 2006, Mueller et al. 2010). The challenge is in finding a way to access those resources for the benefit of humanity.

Until now this has not been economically feasible because of the vast distances and orbital energies separating the bodies in our solar system with the high expense of spaceflight. Gerard K. O'Neill (1989) estimated in the 1980's that an orbiting space colony could become economic only if it had a human population of greater than 10,000 to perform manufacturing tasks. History has since proven that there was little chance of building a space colony if it had to be that large to make a profit. Zubrin (1999) argued in the 1990's that human-tended manufacturing colonies on the Moon or in the asteroid main belt were not practical because the massive energy needed to grow food for the humans and because of the scarcity (in the lunar case) of some elements such as hydrogen, nitrogen and carbon. In their analyses, O'Neill and Zubrin did not include the effects of robotic laborers in lieu of humans to lower the costs, probably because robotic technology was too immature to predict with any confidence. (Similarly, the present study ignores the possible effects of nanotechnology.) Since then, we have seen how advances in robotics and additive manufacturing (3D printing) technologies are game-changing for space colonization. As a result, it has become economically feasible – as we attempt to demonstrate in this paper – to bootstrap a self-sustaining, self-expanding lunar industry that will spread across the solar system at no further expense to the Earth's economy. Another game-changer is the discovery of lunar polar ice providing vast quantities of hydrogen, nitrogen and carbon. The Moon has every element needed for healthy industry. In light of these game-changing advances and discoveries, it is important to reassess the prospects for initiating space industry.

Once successfully bootstrapped, a robotic network can access, process, transport, and utilize the solar system's resources for mankind's benefit. Appropriately designed robots will not have the problems traveling the vast distances of the solar system that humans have, and they can set up the infrastructure that will enable us to follow. Within the first several decades a vital industry could be established on the Moon and in the asteroid belt using technologies that are for the most part only modestly advanced beyond today's state-of-the-art. After that, human outposts, laboratories, and observatories can spring up everywhere between the Kuiper belt and Mercury. It can grow exponentially and provide mankind the ability to do things that today are only dreams. To make this possible very soon, the majority of technology advancement needs to occur in the automation of robotics and in additive manufacturing. Trends in these fields are hopeful, so we think the scenarios we present here are not too optimistic. Therefore, we think the space resource community has real reason to be motivated in its work. We are aiming for the possible, not the fantastical.

This paper roughly assesses how much mass and time are needed on the Moon to reach the "ignition" point of a self-sustaining and expanding industry, and it shows that the launch costs for this mass can be quite low. It does not assess the cost of developing the necessary technologies and of teleoperating them on the Moon until autonomy is achieved. While the mass and time are shown to be quite low, it might be necessary to subdivide the bootstrapping into even smaller, less expensive steps, creatively sharing them between public and private sectors. We leave that business model strategizing to future work. Also, the assessment in this paper is very rough and is intended mainly to organize our thinking on this topic, and to initiate discussion and further study within the space community. A full study will be very complex and require the involvement of a much larger group of contributors. We hope this will raise interest and lead to that more comprehensive effort in the near future.

## Bootstrapping a Solar System Industry

Self-replicating systems have been studied as an innovative method to economically access space resources (Freitas and Gilbreath 1980; Tiesenhausen and Darbro 1980; Freitas and Zachary 1981; Chirikjian 2004).  A 1980 summer study at the NASA Ames Research Center (Freitas and Gilbreath 1980) showed that self-reproducing machines are theoretically possible.  It discussed a straw-man self-replicator of 100 metric tons mass, including 12 tons for paving robots, 4.4 tons for mining robots, and 4 tons for mobile assembly and repair robots, to name a few examples.  Freitas and Zachary (1981) also used the figure of 100 metric tons per replica.  The 1980 study recommended among other things a technology development program for the enabling technologies.  This program has in effect occurred during the past three decades, mostly driven by non-space, commercial industry but also in the past decade by the Constellation project.  As a result, these masses per replica can be reduced.  For example, the excavator masses used in this paper are only 0.35 tons for the first generation "seed" hardware, based on trade studies and our experience with robotic lunar excavators and pavers that were recently developed and field tested (Zacny et al. 2010; Mueller et al. 2009; Mueller and King 2008).

The 1980 study portrayed the seed replicator as a large factory, with warehouse operations, centralized computing, and significant facility construction. Lipson and Malone (2002) showed how Solid Freeform Fabrication technology (or additive manufacturing or 3D printing) could reduce the complexity of a space manufacturing operation.  This would reduce the mass of the first seed replicator.

There are several additional strategies to reduce the launch mass of a seed replicator.  The first is to identify and use only the simplest system capable of replication.  The second is to avoid full "closure".  Closure is the ability to replicate all aspects of the system in space so that nothing further is required from Earth to build replicas.  Nearly full closure is vastly easier to achieve than full closure (Freitas and Gilbreath 1980), because the manufacture of electronics and computer chips requires heavy, high-tech equipment that would be expensive to launch from Earth and would command much of the industry's resources during replication.  However, incomplete closure results in very high launch masses later as the industry grows exponentially, as we show below.  A third strategy, which to our knowledge has not been discussed in the literature, is to begin with a simpler, sub-replicating system and evolve it toward the self-replication capability.  In this strategy, the evolving system might never become a "self-replicator" even after it reaches full closure, because each generation can continue creating something significantly more advanced than itself.  This is the strategy adopted here.

The first hardware sent to the Moon will be high-tech equipment built on Earth.  However, the high launch costs demand that it be mass-limited so it will have insufficient manufacturing capability to replicate itself.  It will construct a set of crude hardware made out of poor materials, so the second generation is actually more primitive and inefficient than the first.  The goal from that point is to initiate a spiral of technological advancement until the Moon achieves its own mature capabilities like Earth's. This evolving approach will provide several benefits.  First, industry on the Moon can develop differently than on Earth.  The environment, the manufacturing materials, the operators (robots versus humans), and the products and target markets are all different.  Allowing it some reasonable time to develop will allow it to evolve an appropriate set of technologies and methods that naturally fit these differences.  Second, the evolving approach supports the development of automation so that industry can then spread far beyond the Moon. The technological spiral will develop the robotic "workers" in parallel with the factories. It will also improve automated manufacturing techniques such as 3D printing. The third and probably most important benefit is the economic one. As we show here, a space economy can grow very rapidly, and it will quickly require massive amounts of electronics and robotics unless there is full closure. The tiny computer chips alone become too expensive to launch within a few decades as the industry grows exponentially, and therefore we will quickly need lithography machines on the Moon to make the computer chips. The evolving approach sends only a small and primitive set of machines as "colonists", and the nascent lunar industry develops over time – but still rapidly – toward the full sophistication that Earth cannot afford to launch. This may seem too far reaching to a reader first exposed to the idea, but the key is the on-going rapid advancement in robotics. After robotic dexterity, machine vision, and autonomy improve for another couple of decades, robots will build lithography machines on the Moon as easily as human workers build them on Earth. This future is not far away, considering the exponential rate of technology development in terrestrial industries. Robotics experts are optimistic that the necessary levels of automation will be developed quickly enough to support the timeline we present here (Moravec, 2003).

So the objective is for the first robotic "colonists" on the Moon to fabricate a set of, say, 1700's-era machines and then to advance them steadily through the equivalent of the 1800's, 1900's, and finally back into the 2000's. We argue that this can be accomplished in just a few decades. There are reasons why this technological spiral will be both easier and faster than when we accomplished it on Earth. First, the majority of the technology does not need to be re-invented. The knowledge will be provided by technologists on Earth. Second, the Earth will provide material support in the early stages. We will send teleoperated robots and complex electronic assemblies prior to achieving closure. On the other hand, there will be new challenges. For example, we must gain experience in the lunar environment to learn how to adapt terrestrial technologies to it.

For comparison, successful bootstrapping on the barren regolith of Earth's continents began with single-celled organisms then fungi and lichens, which created top soil (Kenrick and Crane 1997; Sleep and Bird 2008) so that diverse plants could take hold (Shear 1991), followed by animals, humans, civilization, and increasingly sophisticated industry. To bootstrap industry on the barren regolith of the Moon we do not wish to "begin at the beginning," since that would require nanotechnology as the analog of single-celled life, and that kind of nanotechnology does not yet exist. We also do not wish to "begin at the end", since that would require us to launch a fully networked ecology of robots and industrial assets to the Moon, which is too massive to afford. We wish instead to begin in the middle with large assets like robonauts and 3D printers that require a complex ecology of interdependence, yet without the fully developed ecology. We discuss the plausibility and affordability of this intermediate approach.

For the concept of lunar industry presented here we do not think the term "self-replicator" is appropriate and so we will avoid the term. A self-replicator is by definition self-contained with all of its parts co-located in a complete set. That entire set fabricates a new complete set that is situated in a new location before the next replication cycle begins. This is unlike industry or biology on Earth: neither businesses nor industries are self-replicators. Although biological species are self-replicators, they require a vast number of other species in a highly networked ecology to survive, and the ecology does not operate on a synchronized replication cycle. We think the networked complexity of these examples is the more successful topology for space industry because it is the one that naturally occurs and hence is probably the more efficient and adapted for survival, as well as the more easily bootstrapped through an evolutionary process. We therefore avoid any requirement that the various hardware assets remain together in a closed set, and we allow instead for transportation to develop naturally between multiple, specialized production sites.  Thus, lithography machines to make computer chips can be located in just one laboratory on the Moon, and their products can be transported to other sites for incorporation into robots and machines.   The original facility to house that equipment can be built larger than necessary to allow for expansion and to gain economies of scale.  This aspect of networked complexity is not visible in the modeling presented here, but it will be evident when modeling future expansion of industry beyond the Moon considering that space resources are distributed in "zones" due to solar system formation processes (Hartmann 1985).

## Hardware Elements in Lunar Industry

Each generation in the evolutionary bootstrapping process is characterized by the sophistication of materials and construction methods that the previous generation used to fabricate it; by the diversity of materials that it is able to make in fabricating the next generation; by the degree of robotic autonomy it possesses; and by the quantity of robotics and electronics that must be imported from Earth to make the next generation.  Some characteristics of the evolving generations are summarized notionally in Table 1.

Table 1.  Generations of lunar industry.

| Gen | Human/Robotic Interaction | Artificial Intelligence | Scale of Industry | Materials Manufactured | Source of Electronics |
|---|---|---|---|---|---|
| 1.0 | Teleoperated and/or locally-operated by a human outpost | Insect-like | Imported, small-scale, limited diversity | Gases, water, crude alloys, ceramics, solar cells | Import fully integrated machines |
| 2.0 | Teleoperated | Lizard-like | Crude fabrication, inefficient, but greater throughput than 1.0 | (Same) | Import electronics boxes |
| 2.5 | Teleoperated | Lizard-like | Diversifying processes, especially volatiles and metals | Plastics, rubbers, some chemicals | Fabricate crude components plus import electronics boxes |
| 3.0 | Teleoperated with experiments in autonomy | Lizard-like | Larger, more complex processing plants | Diversify chemicals, Simple fabrics, eventually polymers. | Locally build PC cards, chassis and simple components, but import the chips |
| 4.0 | Closely supervised autonomy with some teleoperation | Mouse-like | Large plants for chemicals, fabrics, metals | Sandwiched and other advanced material processes | Building large assets such as lithography machines |
| 5.0 | Loosely supervised autonomy | Mouse-like | Labs and factories for electronics and robotics. Shipyards to support main belt | Large scale production | Make chips locally. Make bots in situ for export to asteroid belt |
| 6.0 | Nearly full autonomy | Monkey-like | Large-scale, self-supporting industry, exporting industry to asteroid main belt | Makes all necessary materials, increasing sophistication | Makes everything locally, increasing sophistication |
| X.0 | Autonomous robotics pervasive throughout solar system enabling human presence | Human-like | Robust exports/imports through zones of solar system | Material factories specialized by zone of the solar system | Electronics factories in various locations |

The set of assets within each generation is described below. To be conservative, we usually assume that each asset is retired at the end of its generation so that only the more modern assets of the new generation are involved in producing the generation after that (except as noted below for solar cells and robonauts). This is overly conservative, but it allows that hardware failures could disable some new assets that are unable to be repaired while assets from the prior generation continue to operate to take their place.

In Generation ("Gen") 2.5, the use of the decimal place (rather than incrementing to 3.0) indicates that the assets of Gen 2.0 and Gen 2.5 are added cumulatively rather than retiring the Gen 2.0 hardware. We do this because it is necessary to vastly diversify materials manufacturing as quickly as possible, and this is accomplished by creating Gen 2.5 hardware that is no more sophisticated than Gen 2.0 although capable of making different materials.

The technologies needed for mining, chemical processing, and metallurgy are for the most part already existent in Earth's industry. The feasibility of adapting them to the lunar environment has been and is currently being demonstrated by the wide variety of successful space utilization projects described elsewhere in this issue of the journal.

Excavators. The excavators will travel between the digging site and the resource processing site, delivering sufficient lunar regolith each hour to maintain production rates of the other assets. The details of the excavators are unimportant. In our modeling we assumed for specificity that they are small and operate in a swarm. They may also be fitted with paving attachments (Hintze and Quintana 2012).

Chemical plants for volatiles. Dry regolith or an ice/regolith mixture will be deposited into hoppers and then fed into chemical plants. Electrical power for the processes is augmented with thermal power from solar concentrators. One type of chemical plant will be concerned with producing gases and liquids. These fluids will include oxygen, hydrogen, water, hydrocarbons such as methane, and (in more advanced generations) solvents for industrial processes. So far, NASA has developed and field tested only basic oxygen production systems, including hydrogen reduction and carbothermal systems. More complex chemical processes have been developed for terrestrial applications, and by adapting the lessons-learned from the lunar projects it should not be difficult to adapt the other processes to the lunar case, too. For specificity, we have described the chemical plants using particular masses, power levels, and production rates after examining several sources of data. These include analyses of lunar chemical plants (Mendell 1985, Taylor and Carrier 1993) and the actual construction and performance of lunar chemical plants that our team and collaborators have recently field tested on Mauna Kea in 2008 and 2010 (Boucher et al. 2011; Captain et al. 2010; Gustafson et al. 2010a; Gustafson et al. 2010b; Muscatello et al. 2009). The specifics are not too important as we will vary these numbers over wide ranges to demonstrate general feasibility of the bootstrapping process. Gen. 3.0 and subsequent will have larger throughputs than the earlier generations, and they will gain from economies of scale by building much larger chemical plants rather than reproducing a large number of smaller plants (Lieberman 1987; Gallagher et al. 2005). However, to be conservative we ignore the economies of scale and instead describe the chemical plants as though they were units identical to the originals. These represent "units" of chemical processing capability in larger plants rather than standalone assets.

Chemical plants for solids. Chemical plants are also needed to produce plastics and rubbers from the lunar polar ice. This is possible because we now know that the ice contains large quantities of carbon molecules (CO, CO2,…) as well as nitrogen-bearing and hydrogen-bearing molecules (Colaprete et al. 2010; Gladstone et al. 2010).  These materials may be needed for gaskets, seals, and insulators, for example. Later diversity will introduce sheet materials, fabrics, and layered and complex materials.  Again, economies of scale are ignored in the model to be conservative.

Metal and Ceramics Refinery.  It will be crucial to manufacture metals and metal-oxide ceramics and to improve the properties of the various alloys with subsequent generations.  Processes to do this from lunar soil have been described (Rao et al. 1979; Jarrett et al. 1980; Sargent and Derby 1982; Lewis et al. 1988; Stefanescu et al. 1988; Landis 2007; Lu and Reddy 2008), and some development work is on-going by our colleagues and collaborators.  Notionally, the early generations in our model will produce crude "mongrel alloys" by electrowinning or other methods.  Hardware constructed from those alloys will need to be massive to add strength to make up for their poor mechanical properties.  (This will be partially offset by the reduced forces in low lunar gravity.)  Subsequent generations of metal refineries will add processes and material streams to improve the mechanics of the materials.  Oxygen and other gases produced by metal refining will be sent to the chemistry plants.  Electrical power is augmented with thermal power collected by solar concentrators.

Manufacturing.  Additive manufacturing will have two forms:  3D printers that make parts that are small enough to fit inside the printer, and larger units that move about robotically and add material onto large structures external to themselves.  The printers may eventually have multiple material streams including metals and plastics to make complex assemblies in a single pass.  However, the earlier generations will require import from Earth of the most complex assemblies, such as electronics packages and the assembly robots.   Furthermore, "appropriate technology" will mandate the design of simpler assets that can function without too many complex or miniaturized components, simplifying their manufacture and reducing imports.  To achieve the final generations, the additive manufacturing technologies require advancement beyond the current state of the art.  However, gains are being made rapidly and it is very likely the advancements will support the bootstrapping strategy presented here.

Solar Cell Manufacturer.  Power will be provided mainly by solar cells.  Ignatiev et al. (2001) and Freundlich et al. (2005) have shown how these may be manufactured on the Moon even in the earlier, more primitive generations of lunar industry.  For specificity, we have described the mass, power, and throughput of the solar cell manufacturers as per those earlier studies.  We show that these devices in the first and subsequent generations produce far more available power than needed by the following generation.  This excess power capacity grows exponentially.  We assume that solar cells are added cumulatively from one generation to the next.  Failure of solar cells by radiation damage and micrometeoroids has not been modeled explicitly, but can be deducted from the exponentially growing excess.

Power Station.  In the first generation, a power station is included in the mass of hardware shipped to the Moon.  This station includes power conditioning, docking stations, and cabling to manage and distribute the solar power.   It might also include a small nuclear reactor to support human presence and as a backup system to support re-bootstrapping in case of system failure.

Robonauts.  Robotic astronauts, or "robonauts", will perform the assembly and maintenance tasks.  The name is borrowed from a particular robot developed by General Motors and the NASA Johnson Space Center, with the assumption that robonauts in future lunar industry will be the direct descendants of the current ones.  The number of robonauts must grow rapidly as the industry itself grows.  At first the robonauts are imported from Earth.  To keep the strategy slightly more economical, they are not retired with each subsequent generation.  Beginning with the third generation their structural components are made on the Moon, while Earth continues to send their cameras, computers, motors, and sensors.  Eventually they are made completely on the Moon.

At first the robonauts will be teleoperated from Earth.  The approximately 2.5 second round-trip communication time delay can be managed even for fine motor tasks, such as screwing parts together by hand, by having a teleoperator on Earth interact with a virtual world that models the robonaut and its environment rather than interact with the reality itself.  The robonauts on the Moon will mimic the behaviors they observe in the virtual reality as closely as possible using existing levels of robotic autonomy.  Resynchronization will occur in the virtual world using a rubric designed to prevent operator confusion.  Similar schemes are being developed for telesurgery with large communication latency (Haidegger and Benyó 2003).  This will make teleoperation manageable for lunar operations, but it will require a growing and expensive workforce of teleoperators on Earth plus sufficient communications bandwidth, and it will not be extensible to the asteroid main belt or beyond.  Therefore, with each generation, progress will be made toward full autonomy.

Table 1 describes the autonomy in terms developed by Hans Moravec (Moravec 1999; Moravec 2003).  Moravec's "insect" level is when robots perform simple pre-programmed responses to sensor inputs.  Many machines operate at the insect level today.  The "lizard" level is when robots identify objects functionally to guide their motor tasks.  Lizard-level robotics is already appearing in laboratories on Earth and is making steady progress toward greater capability.  "Mouse" level is when the robots learn and improve the performance of their tasks through simple positive and negative feedback.  This is important because human industry is only adapted to terrestrial conditions, but learning robots can adapt it to the multitude of environments they will experience in the solar system.  "Monkey" level is when the robots maintain a mental model of the world including other agents.  This provides them with insight into the intent of other agents as well as foresight.  "Human" level is when the robots have the mental ability to reason abstractly, generalizing from specific learning situations to a broader class of applications, and thus to make decisions in the face of the unexpected.  These higher levels of robotics will be necessary in the distant future when, for example, a robotic construction crew is building a science lab on Pluto, many hours of time delay away from human help.  Extrapolating the computing speed of small, inexpensive microprocessors that are commercially available, we expect by the year 2023 they will reach the speed Moravec predicted as necessary to support human-level robotics.  Even if Moore's Law ended today, that computer power is easily achieved by paralleling inexpensive microprocessors and by other advances planned by computer chip manufacturers (Gargini 2005).  On-going advances in robotic software and artificial intelligence present a very optimistic picture that these levels of robotics will be achieved as Moravec predicted, with lizard-level occurring by 2020, mouse-level by 2030, monkey-level by 2040, and human-level by 2050.  Only mouse-level is needed by the end of bootstrapping on the Moon, but depending on how fast the strategy is carried out the robotics sent to the asteroid belt may be at the monkey-level or higher.

<u>Electronics Manufacturing.</u>  In the baseline model, when Gen. 2.5 is fabricated its assets include some electronics manufacturing machines.  Those machines themselves are built with electronics imported from Earth, and they are capable of making only the crudest and simplest of electronics components such as resistors and capacitors, which will not be miniaturized or efficient. Gen. 3.0 and subsequent possess a greater diversity of electronics manufacturing machines with increasing sophistication. By Gen. 5.0 we aim to have basic lithography machines on the Moon, built using computer chips sent from Earth, so that by Gen. 5.0 all computer chips can be made in space. The early computer chips will lack the transistor density of chips made on Earth, but they will be adequate for "appropriate technology" in space. Later generations (not modeled here) continue spiraling the sophistication of space industry so that eventually the lithography machines and computer chips match the best of Earth's.

## Modeling the Bootstrapping

The process of bootstrapping was modeled in spreadsheet form using equations for mass balance, energy balance, quantities of the various assets, and production times. The model follows the hardware through six generations of increasing complexity until full independence from the terrestrial economy is achieved. The parameters of the baseline model were determined as described above for the various assets and were varied over wide ranges to determine that the model produces robust and reasonable conclusions despite its simplicity. Varying the parameters also identifies system dependencies and sensitivities. Identifying these provides further indication of the probability of success in economical bootstrapping; if it turns out that the only sensitive parameters are easily adjusted to put them into more manageable ranges, or if alternative hardware configurations can remove those sensitivities entirely, then economical bootstrapping should be easily achieved. The parameters of the model in Gen. 1.0 are shown in Table 2.

Table 2. Baseline values for Generation 1.0 in Bootstrapping Model.

| Asset | Qty. per set | Mass minus Electronics (kg) | Mass of Electronics (kg) | Power (kW) | Feedstock Input (kg/hr) | Product Output (kg/hr) |
|---|---|---|---|---|---|---|
| Power Distrib & Backup | 1 | 2000 | – | – | – | – |
| Excavators (swarming) | 5 | 70 | 19 | 0.30 | 20 | – |
| Chem Plant 1 – Gases | 1 | 733 | 30 | 5.58 | 4 | 1.8 |
| Chem Plant 2 – Solids | 1 | 733 | 30 | 5.58 | 10 | 1.0 |
| Metals Refinery | 1 | 1019 | 19 | 10.00 | 20 | 3.15 |
| Solar Cell Manufacturer | 1 | 169 | 19 | 0.50 | 0.3 | – |
| 3D Printer 1 – Small parts | 4 | 169 | 19 | 5.00 | 0.5 | 0.5 |
| 3D Printer 2 – Large parts | 4 | 300 | 19 | 5.00 | 0.5 | 0.5 |
| Robonaut assemblers | 3 | 135 | 15 | 0.40 | – | – |
| **Total per Set** | | **~7.7 MT launched to Moon** | | **64.36 kW** | **20 kg regolith/hr** | **4 kg parts/hr** |

Subsequent generations are modeled very simply as extrapolations from the first one using a "crudeness factor" that tells how much more massive they are due to the use of mongrel alloys and other poor materials produced from the regolith. Thus, Gen. 2.0 and Gen 2.5 have a crudeness factor of 2.5, meaning they are 2.5 times as massive as Gen. 1.0 and thus take longer to manufacture. Gen 3.0 has a crudeness factor of 1.5, but Gen. 4.0 and subsequent have a crudeness factor of 1.0.

The quantity of electronics fabricated on the Moon also evolves with the generations.  For Gen. 3.0, 4.0, 5.0 and 6.0, the targets are to make 90%, 95%, 99%, and 100% of the electronics on the Moon, respectively.  If these targets cannot be met, then the overall exponential growth of lunar industry can be slowed down accordingly to keep it economical.

Another model parameter is the operation time per lunation.  If the solar cells are located on a "peak of eternal light" near the poles and are actuated to follow the sun, they might obtain enough power for the industry to operate through 70% of the lunation.  More equatorially, they would support only 40% operation.

A flow diagram of the model is provided in Fig. 1.  Because each generation feeds information to the generations both before and after it, the model is recursive and must be iterated for consistency each time any parameter is changed.  The numbers of excavators, solar cell manufacturers, and fluids chemical plants have not been optimized but instead set to values much higher than needed.  Excavators have many other jobs in building landing pads, stabilizing roads, and preparing surfaces for building construction; solar cell pavers make excess solar cells for conservatism; and fluids chemical plants produce large quantities of propellants or other consumables for transport vehicles traveling to and from Earth and possibly for a human-tended outpost.  Human presence is highly desirable (but not mandatory) in the early parts of bootstrapping.  Spudis and Lavoie (2010) showed how a human outpost is very affordable when based on the use of lunar resources.  It is assumed that the consumables are provided to an outpost or to commercial businesses that transport them profitably to low Earth orbit for other space operations, so construction of storage tanks has not been included in the production budget.

Additionally, in Gen. 3.0, 80 metric tons (MT) of construction equipment are fabricated (not shown on Fig. 1 for simplicity).  The production rate of this equipment is subtracted out from the total production rate and hence reduces the number of basic sets of assets that can be fabricated by that generation.  The constructed equipment is needed in Gen. 4.0 so that dust-free laboratories can be built where the more sophisticated electronics manufacturing including lithography machines will be housed.  In Gen. 4.0, a total of 10 MT of metals are set aside as reinforcement for the fabrication of those buildings out of regolith (also not shown on Fig. 1 for simplicity).  In Gen. 5.0, materials are stockpiled for the construction of a fleet of spacecraft to export the industry to the asteroid main belt (also not shown on Fig. 1).  This fleet transports 72 MT of industrial equipment and robonauts as the seed for Main Belt industry.  Each spacecraft is assumed to have 20 MT dry mass and to carry 12 MT payload, so there will be six such vehicles in the fleet for a total of 120 MT materials set aside by Gen. 5.0.  The delta-v to Ceres is 9.5 km/s, and the specific impulse of a hydrogen/oxygen engine is about 455 s.  The required propellant mass for this fleet is therefore about 1400 MT, which is just 2.8% of fluids produced by the six generations of lunar industry as modeled here by the baseline case.  This seems very feasible.

## Results of the Modeling

The modeling shows that bootstrapping a lunar industry according to this strategy may be affordable over a wide and accessible region of parameter space.

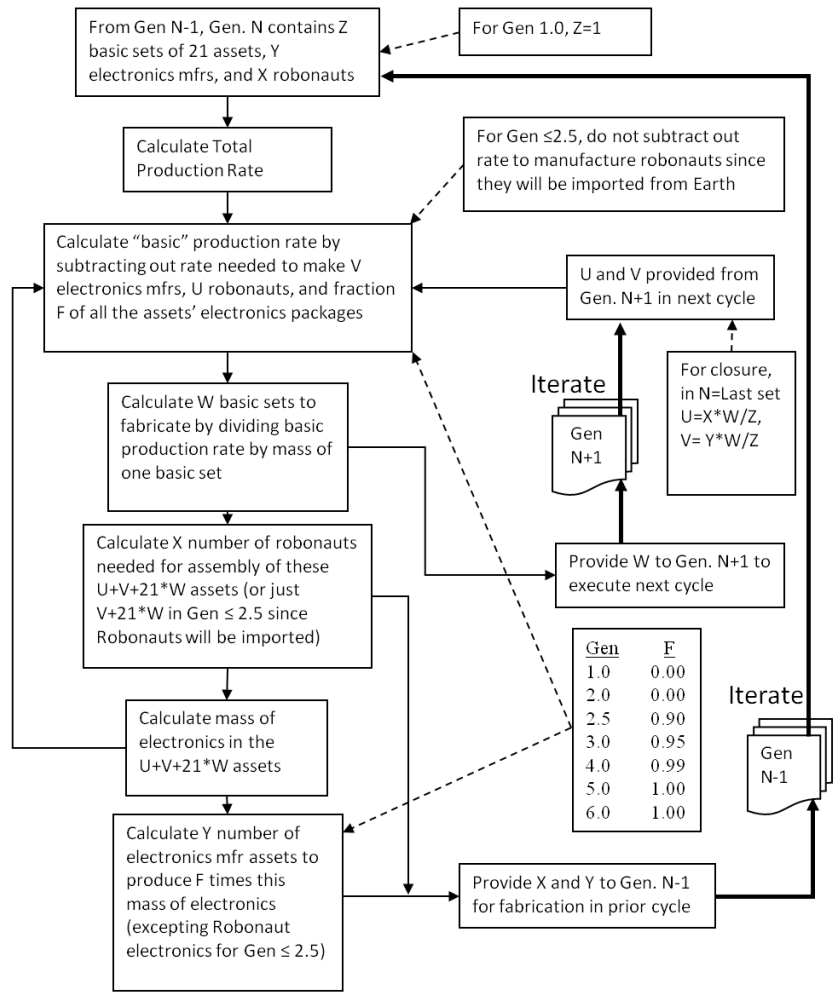

Figure 1. Flow diagram of model, showing Gen. N manufacturing Gen N+1. Heavy lines represent iterated relationships between generations. Omitted details are described in the text.

Growth Rate of Space Industry

The model predicts that the lunar industry can grow exponentially, as shown for two cases in Figs. 2 and 3. The growth is dramatic and rapid, but it is reasonable considering that each basic set of assets creates two or fewer new sets per year (plus the robonauts and other set-asides such as construction of facilities and spacecraft in later generations). These cases assume a 70% solar power duty cycle and generations spaced in 2-year intervals. The manufacturing rate is assumed to be 0.5 kg/hr for each additive manufacturing unit. This is a high estimate and will be discussed below. The first case uses this rate to reproduce hardware as quickly as possible and thus achieve the largest space industry possible year-by-year. The second case voluntarily pauses the manufacturing within each generation to allow for other tasks, such as technological advancement, or manufacturing experiments, or for slow robonauts to catch up assembling parts. This reduces the time-averaged manufacturing rate to half its maximum, which we find is the minimum necessary to meet the survival and growth goals until full autonomy is achieved.

This results in a plateau of asset mass near 100 MT, delaying the exponential growth until after full closure and thus minimizing the launch costs.  It might be significant in Fig. 4 that the mass plateau hovers around 100 MT because this is the same mass estimated in the NASA/Ames summer study of 1980 (Freitas and Gilbreath 1980) for a single "seed" replicator.  Thus, our study agrees that this is the correct order of magnitude for industry to ignite while being just shy of full closure.   However, by evolving toward this mass over several generations, only a small fraction of it needs to be launched from Earth (as shown below) and in the end it achieves full closure, which the earlier non-evolving strategy did not attempt.

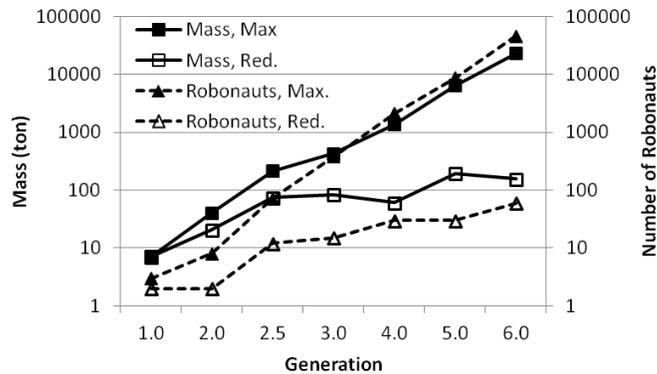

Figure 2.  Growth of lunar industry by generations in 2-year intervals.  Connecting lines are a guide to the eye.  Solid markers – Case with maximum manufacturing rate, demonstrating exponential growth.  Open markers – case with manufacturing rate reduced by half.  Solid lines – mass of assets, including both hardware brought from Earth and hardware built on the Moon.  Dashed lines – number of robonauts.

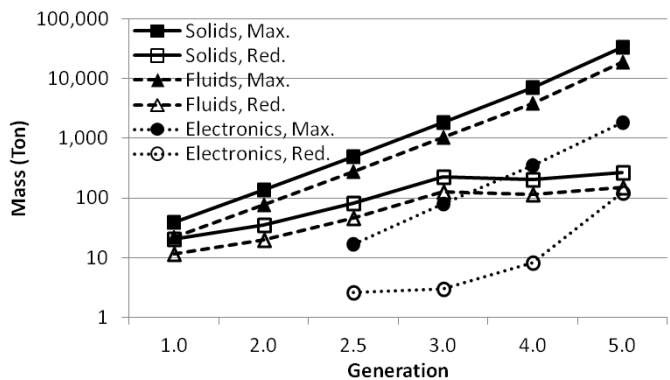

Figure 3.  Production of materials and parts by each generation in 2-year intervals.  "Max" and "Red." refer to maximum and reduced manufacturing rates.  Solids includes both plastics/rubbers and metals, but not electronics.

Launch Mass

The mass of hardware that must be launched to the Moon throughout the process is shown in Fig. 4 for the maximum and reduced manufacturing rate cases.  In the baseline case with maximum manufacturing rate, a total of 41 MT is launched to the Moon.  In the reduced rate case, only 12 MT is launched to the Moon but the scale of industry is two orders of magnitude smaller so the benefits of the industry are relatively delayed by eight years.  Figure 4 also shows how much mass would be needed from Earth year-by-year if full closure is not achieved:  no electronics and/or no computer chips manufactured on the Moon.  Those masses grow exponentially, and this demonstrates the need to achieve full closure.

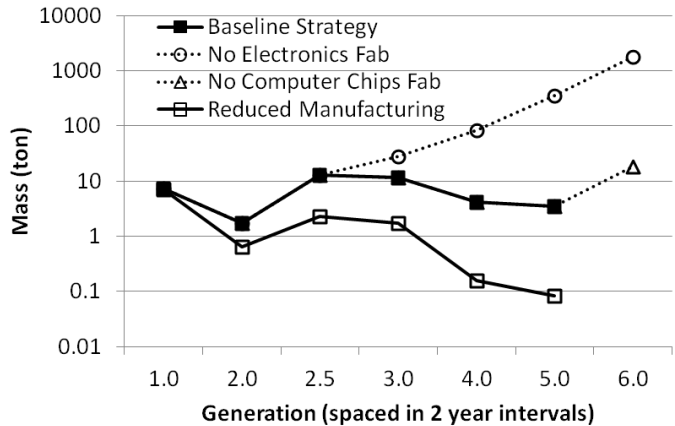

Figure 4.  Mass launched to Moon for each generation spaced in 2-year intervals.  Lines are a guide to the eye.  Solid squares – baseline case operating at maximum manufacturing rate.  Variants:  Open squares – manufacturing rate reduced to half of capacity.   Open circles – no electronics or computer chips made on the Moon.  Open triangles – no computer chips made on the Moon.

Manufacture of Solar Cells

Power needs and power availability are shown in Fig. 5.  In Gen. 1.0, the power is from an asset launched directly from Earth.  In subsequent generations it is from solar cells fabricated on the Moon.  To be conservative, we scaled the fabrication of solar cells so that the available solar power would be much greater than required throughout the process.  This margin allows that solar cell fabrication may not be as efficient as predicted, and it provides excess power for other purposes.

Lunations and Solar Availability

The effects of reducing the solar duty cycle (i.e., solar power availability per lunation) from 70% to 40% are shown in Figs. 6 and 7.  They indicate the importance of the duty cycle, because reducing it by less than a factor of two decreases the net growth by more than an order of magnitude.

Printer Speed

Other than the initial launch costs, we do not care about the hardware mass as an independent parameter; we only care about mass in comparison with manufacturing speed, because their ratio predicts how long it takes the system to reproduce.  Modeling identifies this ratio as a critical parameter.  Currently, commercially-available additive manufacturing prints either metals or plastics at 0.10 to 0.15 kg/h.  Although the modeling shows this can successfully bootstrap the lunar industry with the masses we have estimated here, it probably needs to be higher for practical reasons.  We do not think we can lower hardware masses more than an order of magnitude.  Therefore, with such a low printing speed, each generation will be required to function for as long as 10 years to produce the next generation as shown in Fig. 8, and during

Preprint. To appear in Journal of Aerospace Engineering.

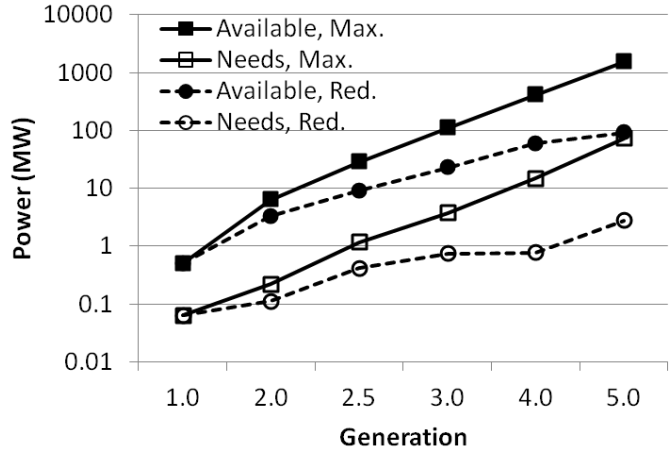

Figure 5. Power needs (dashed lines) and power generated (solid lines). The cases with maximum manufacturing rate (solid markers) and reduced manufacturing rate (open markers) both have vastly excess power capacity.

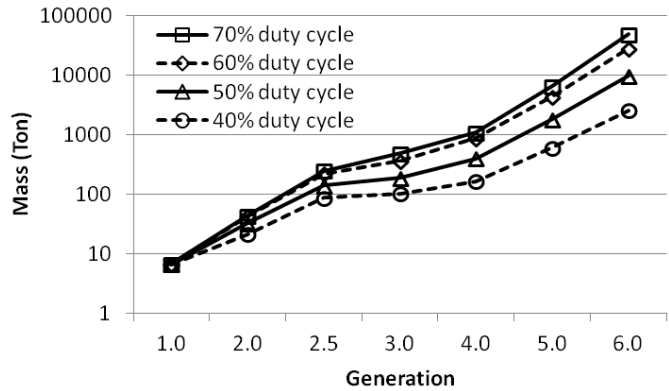

Figure 6. Mass of assets per generation for various solar power duty cycles. In these cases, production rate is 1 kg/h and each generation lasts 2 years.

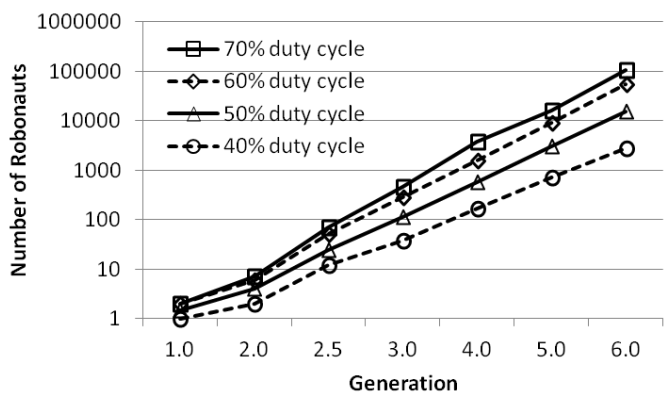

Figure 7. Number of Robonauts per generation for various solar power duty cycles. In these cases, production rate is 1 kg/h and each generation lasts 2 years.

that time there will be hardware failures and growing labor costs for the teleoperators on Earth. Adding more printers to the hardware set will increase the total printing rate, but this is a self-limiting strategy because each printer has its own mass that must be reproduced for the next generation. In the limit of large numbers of printers the mass of all other assets becomes insignificant and the reproduction time approaches the "specific reproduction" of the printer: its mass divided by its production rate averaged over a lunation. For a 300 kg (large) printer operating at 0.1 kg/h for 70% of a lunation, the specific reproduction is about 3 months. However, with a crudeness factor of 2.5, it will take 7.5 months for Gen. 1.0 to make Gen. 2.0, and that limit can be achieved only with a huge number of printers with extremely large launch mass. Figure 8 shows these reproduction times predicted by the model for Gen. 1.0 as a function of increasing numbers of printers at various production rates. Only limited benefit is obtained by reducing the printer mass in half, and for smaller (more reasonable) numbers of printers per set the printer mass is not a significant parameter. (The total mass of assets must be reduced to significantly reduce the reproduction time.) Therefore, apart from dramatic reductions in total asset mass, we can achieve acceptably small reproduction times on the order of a few months only by increasing the throughput of the manufacturing units by a factor of four or greater. This might be easily achieved over the next few years with some technology investment (Lott, 2009). However, it is likely we will need to use a combination of manufacturing techniques to get a sufficiently high manufacturing rate. Since huge numbers of identical parts will be produced, it makes sense to use the additive manufacturing units to print the injection moulds and then to cast the parts in aluminum or iron at a much higher rate. Robonauts will be capable of grinding, polishing or otherwise machining the parts after removal from the moulds. The use of casting, plus other manufacturing processes where appropriate, should easily achieve a 1 kg/h rate.

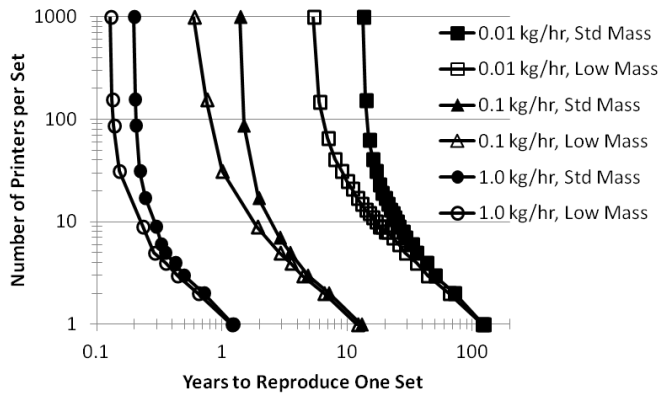

Fig. 8. Numbers of printers per set vs. reproduction time in years for Gen. 1.0 (crudeness factor 2.5 for the next generation). The standard mass ("Std. Mass") cases use equal numbers of the smaller 169 kg printer and the larger 519 kg unit (or one extra of the smaller size for odd numbers of printers). The "Low Mass" cases are with these masses divided by two.

The maximum production of a generation is the printer rate times the solar duty cycle times the length of the generation. In the baseline case with 70% duty cycle, 0.4 kg/h printer speeds, and 2-year generations, maximum production is about 6 MT. One scheme to offset slower printers is to increase the length of the generations to keep total production constant. Figures 9 and 10 show the mass of assets and the number of robonauts, respectively, produced by each generation when the maximum production is 6 MT at various printer speeds. It is interesting to note that by the end of bootstrapping there are more robonauts for the high printer speed cases but more assets for the low printer speed cases. This is because with higher printer speeds the generations are very short and more robonauts must be manufactured in one generation to accomplish the assembly of the following generation in its own short window. Diverting resources to the manufacture of robonauts takes them away from the manufacture of basic sets of assets. The relationships are non-monotonic, as Fig. 11 illustrates.

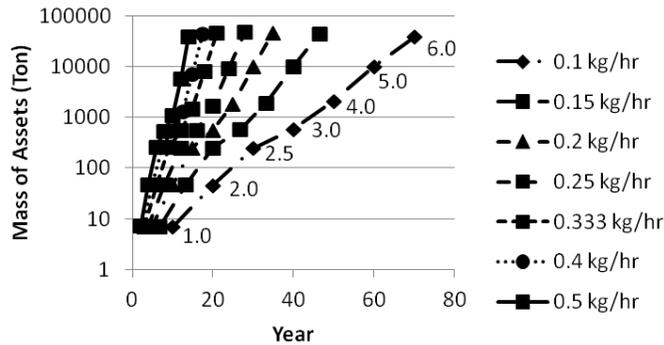

Figure 9. Mass of assets in each generation by year of its completion for various manufacturing speeds. The generation numbering labels are annoted on the chart. In these cases, the maximum production per generation is held constant at 6 MT.

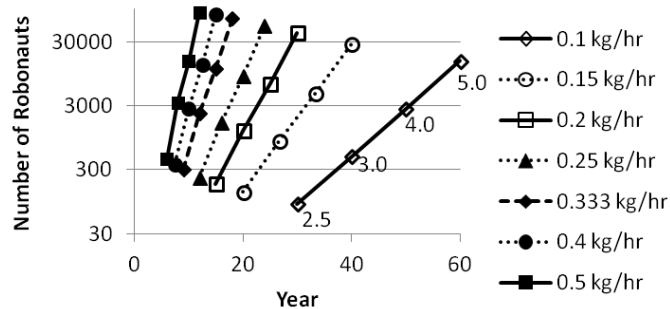

Figure 10. Number of robonauts fabricated by each generation (non-cumulative) versus year of completion of each generation, for the same cases shown in Fig. 9. The numbering labels of the generations are annoted on the chart. No robonauts were made by the first two generations.

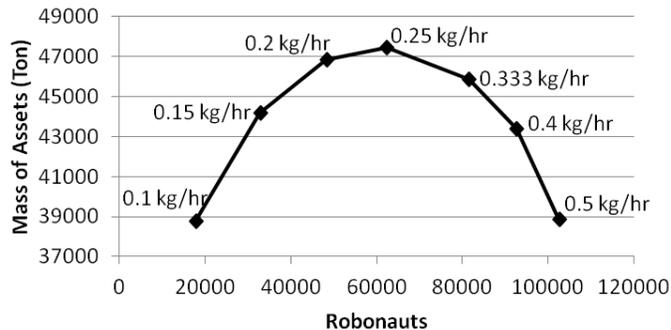

Figure 11. Mass of assets versus number of robonauts manufactured for Gen. 6.0 for the same cases shown in Figs. 9 and 10.

Robonaut Productivity

In the model, the robonaut productivity is described by the "Robonaut Weeks Per Asset" (RWPA) metric, defined as the number of robonaut work weeks that are required to assemble one "average" asset.  The model calculates how many assets will be constructed by a generation, with each excavator, robonaut, unit of chemical plant, etc., counting as one asset. It multiplies the total count by the RWPA and then divides by the number of available work weeks to calculate how many robonauts will be needed to finish assembly on time in that generation. In the nominal case, the RWPA for successive generations is 4, 4, 6, 7, 7 and 8, increasing because the assemblies become more sophisticated as "appropriate technology" evolves into "mature technology" and presumably takes longer to assemble.  To be conservative it is assumed that the increasing autonomy and improving dexterity of robotics cannot fully compensate for this so the RWPA must increase.  The dependence on robonaut productivity is tested by increasing RWPA by a factor of three for each generation, respectively, to simulate slower robonauts, or by dividing it by a factor of three to simulate faster robonauts, while keeping the generation period constant at 2 years.  These cases are for printer speeds of 0.4 kg/h operating at maximum rate, and for a solar duty cycle of 70%. Figure 12 shows that for larger RWPA the number of robonauts must increase in the earlier generations, because the minimum asset set must be completed for the industry to survive no matter how slow the robonauts are.  In later generations, for larger RWPA the number of robonauts is the same as in the nominal case, but they manufacture fewer assets.  Figure 13 shows how this affects growth of the lunar industry.  Figure 14 shows the effects upon launch mass.  Faster robonauts lowers the total launch mass to about 12 MT.  For slower robonauts, more of them are required in the earlier generations to achieve minimal survival of the industry, and since they are being shipped from Earth in the early stages, the total launch mass increases dramatically to about 83 MT. However, an alternative strategy would be to stretch out the generation time longer than 2 years.  Thus, no matter how large the RWPA, the production of parts can occur at a leisurely pace governed by slow assembly speed of the robonauts, and launch mass can be maintained at or below nominal values by stretching out the bootstrapping process another 6 to 8 years.

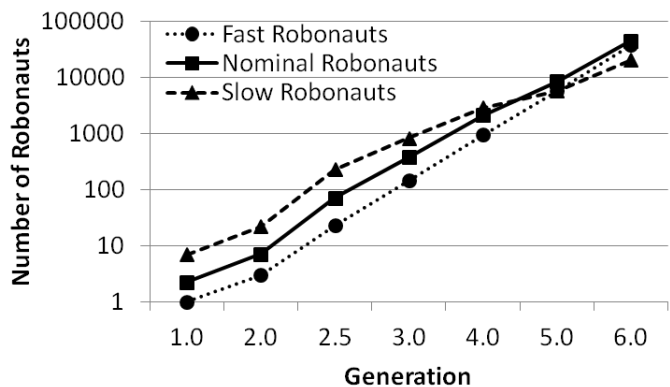

Figure 12.  Number of Robonauts per generation for various robonaut productivity rates.  Cases with fast and slow robonauts assume RWPA of 1/3 and 3 times the nominal case, respectively.

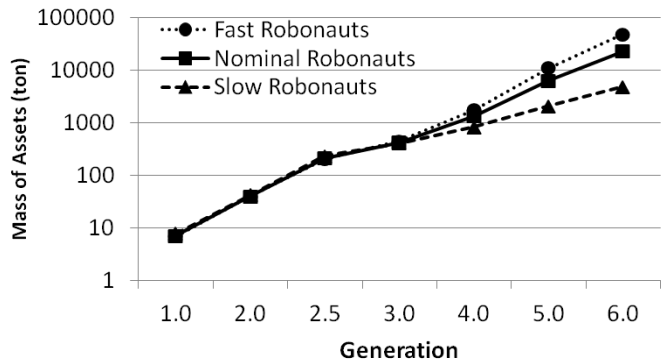

Figure 13.  Growth of asset mass for various robonaut productivity rates.  Cases with fast and slow robonauts assume RWPA of 1/3 and 3 times the nominal case, respectively.

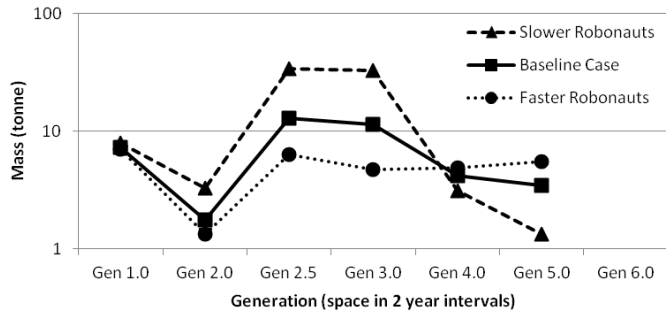

Figure 14.  Required launch mass for each generation in two-year intervals with various robonaut productivity rates.  Cases with fast and slow robonauts assume RWPA of 1/3 and 3 times the nominal case, respectively.

## Discussion

The details of the modeling results are not as important as the general picture it paints:  that *bootstrapping space industry can be achieved in a very short time, for relatively little cost, beginning immediately*.  What makes this possible now is that robotics with artificial intelligence is advancing at a pace to make machines as capable as humans at complex manufacturing tasks within a few decades.  The automation is already sufficient to support teleoperated assembly of parts that are manufactured on the Moon.

Launch costs are one of the main barriers to activity in space.  They are typically on the order of $10,000 (US dollars) per kilogram to low earth orbit (LEO) (Federal Aviation Administration 2009; Perez 2011).  It is difficult to estimate the cost of transporting hardware to the lunar surface because that capability has never existed commercially.  The gear ratio is about 4:1 (Rapp 2010), meaning that for every one kilogram landed near the lunar poles approximately four kilograms must be launched from Earth to LEO.  Assessing just the LEO costs with this gear ratio, a 100 MT seed replicator on the lunar surface would have required $4 billion in launch costs to LEO.  Reducing the lunar surface mass to 12 MT via the evolving approach presented here reduces the LEO launch costs to about $0.5 billion.  Furthermore, SpaceX is expecting about $1500/kg to $2400/kg launch cost to LEO in the near future with its Falcon 9 Heavy launch vehicle (Space Exploration Technologies Corp. 2012).  This brings the LEO launch costs even lower to about 2-3% of the original figure.  Future studies may assess cost of transportation to the lunar surface and the cost of labor to develop and teleoperate lunar industry, but clearly the prospects for initiating the industry are much better now than they were in 1980.

This robotic industry in space leads to a grand vision.  After the industry becomes self-supporting it can be sent to other parts of the solar system.  The asteroid belt has everything necessary for it:  water, carbon, silicates, metals, oxygen, solar energy (with much larger collecting arrays), etc.  The ices in the lunar poles are a limited resource so it will be important to move the center of industry to the asteroids as quickly as possible.  There, the billion-fold greater resources could allow the industry to expand exponentially until it dwarfs that of the entire Earth within just a few decades.  Continued advances in artificial intelligence will be needed to control and manage such a large industry.  The United States economy uses $10^{20}$ J of energy per year including fossil fuels, nuclear, and renewables (Department of Energy 2010).  Figure 3 shows Gen. 5.0 (at 70% duty cycle) using $10^{15}$ J of energy per year.  Multiplying this by a factor of 3 per year, it would exceed the energy usage of the US within 11 more years.  After 12 more years it would exceed the US economy by a factor of a million.  After another decade it would exceed the US economy by a factor of a billion.  Somewhere within that brief period of time, humanity will have gained the ability to do everything we can dream of doing in space.

Robotic space industry will also bring great dividends back to Earth.  For example, it can create space beamed solar power (SBSP) satellites in Earth orbit.  Commonly it is explained that SBSP is not competitive with other energy sources due to launch costs.  Robotic space industry can eliminate not only the launch costs but also the construction costs, turning it into an essentially free, clean, and highly scalable energy source.

The modeling also indicates a significant national security risk.  On Earth, the industry of a nation is limited by its resources including real estate, energy, ores, and the education and size of its labor pool.  In a robotic industry occupying a solar system, the resources and real estate are a billion times greater. Education and talents are learned once then transmitted electronically to all robotic laborers, which are mass produced by the industry itself.  Until this industry begins to feel the limits of the entire solar system, it can grow exponentially.  If any nation initiates and controls such an industry first, then it will have a perpetual lead in industrial power over any other nation that initiates the same capability second.

Space industry will fundamentally change the status of humanity in the solar system.  Humans are a species adapted to living at the top of a food chain, and we need something analogous to the food chain in space to process its resources for us.  That would free us from activities related to merely surviving in space, enabling a greater focus on activities that are uniquely human.  When we spread robotic industry throughout the solar system, it will become the analog of Earth's biosphere and food chain.  We might call it a *robotosphere*.  The interrelationship of robots in space may be studied by robo-ecologists learning how to optimize the robotosphere in support of humanity's goals.  We will want to design it for aesthetics as much as for functionality.  It can work under the direction of human artists and architects to fill outposts and cities throughout the solar system with works of beauty.  It can participate with biological life in terraforming planets and moons in this solar system and beyond.  This robotic/biological ecosphere will become the focus of lifelong study for future generations of scientists.

Preprint. To appear in Journal of Aerospace Engineering.

The analogy to biological ecology is not perfect. On Earth, most life forms get their resources in one location. Most that migrate do so seasonally and get everything they need in one location at a time. In the solar system, the robots will probably not get everything they need in any one zone. Metals are mostly available in the asteroid belt, while volatiles are mostly in the outer solar system, for example. To fulfill their potential the robots will need to set up logistics chains, transporting resources between the various zones. In a sense the robotosphere will function as a single organism. By sending seed ships to industrialize other solar systems, it produces offspring. Riding along as endosymbionts of this organism is one way for man to travel between the stars and extend human culture throughout the Milky Way.

## Conclusion

We are not yet able to model future space industry in much detail, but we can explore some of its main features in this sort of simple model. The modeling looks very optimistic as we vary its parameters to study their relationships, and it indicates that bootstrapping a space-based, robotic, self-sustaining industry is eminently feasible. If we begin working on it today, a vibrant solar system economy will occur within our children's lifetimes or possibly within our own. All of the benefits of its billion-fold industrial power will be at humanity's service for no cost beyond the initial investment of 12 to 41 MT of hardware landed on the Moon (per Fig. 2), plus the cost of a modest robotics and manufacturing development program leading up to that and then the labor to teleoperate the systems until they become autonomous. With its self-sustaining industrial might we can then provide resources back to Earth, clean up the Earth, terraform Mars, build space colonies, support science and the humanities in a well-endowed institute located in space, and send replicas of our robotic industry to other solar systems where it will do all these same things in advance of our arrival.

Preprint. To appear in Journal of Aerospace Engineering.